\documentclass[
journal=jacsat, 
manuscript=article]{achemso}

\usepackage{amsmath}
\usepackage[utf8]{inputenc}
\usepackage{graphicx}
\usepackage[labelfont=bf]{caption}
\usepackage[T1]{fontenc}

\DeclareFontFamily{OMS}{oasy}{\skewchar\font48 }
\DeclareFontShape{OMS}{oasy}{m}{n}{%
         <-5.5> oasy5     <5.5-6.5> oasy6
      <6.5-7.5> oasy7     <7.5-8.5> oasy8
      <8.5-9.5> oasy9     <9.5->  oasy10
      }{}
\DeclareFontShape{OMS}{oasy}{b}{n}{%
       <-6> oabsy5
      <6-8> oabsy7
      <8->  oabsy10
      }{}
\DeclareSymbolFont{oasy}{OMS}{oasy}{m}{n}
\SetSymbolFont{oasy}{bold}{OMS}{oasy}{b}{n}

\DeclareMathSymbol{\smallleftarrow}     {\mathrel}{oasy}{"20}
\DeclareMathSymbol{\smallrightarrow}    {\mathrel}{oasy}{"21}
\DeclareMathSymbol{\smallleftrightarrow}{\mathrel}{oasy}{"24}

\newcommand{\tensor}[1]{\overset{\scriptscriptstyle\smallleftrightarrow}{#1}}
\renewcommand{\boldsymbol}{\mathbf}

\makeatletter
\def\acs@author@fnsymbol#1{}
\makeatother

\email{*markas@dtu.dk}
 
\author{Mark Kamper Svendsen$^{\mathrm{1},\dagger,*}$}
\altaffiliation{$^\dagger$These authors contributed equally to the manuscript}
\affiliation
{$^1$CAMD, Department of Physics, Technical University of Denmark, 2800 Kgs. Lyngby, Denmark}
\email{*markas@dtu.dk}

\author{Yaniv Kurman$^{\mathrm{2},\dagger}$}
\affiliation[Techion]
{$^2$Department of Electrical Engineering, Technion, Israel Institute of Technology, 32000 Haifa, Israel}

\author{Peter Schmidt$^\mathrm{3}$}
\affiliation[ICFO]
{$^3$ICFO-Institut de Ciencies Fotoniques, The Barcelona Institute of Science and Technology, 08860 Castelldefels (Barcelona), Spain}

\author{Frank Koppens$^\mathrm{3}$}
\affiliation[ICFO]
{$^3$ICFO-Institut de Ciencies Fotoniques, The Barcelona Institute of Science and Technology, 08860 Castelldefels (Barcelona), Spain}

\author{Ido Kaminer$^\mathrm{2}$}
\affiliation[Techion]
{$^2$Department of Electrical Engineering, Technion, Israel Institute of Technology, 32000 Haifa, Israel}

\author{Kristian S. Thygesen$^\mathrm{4}$}
\affiliation[DTU]
{$^4$CAMD and Center for Nanostructured Graphene (CNG), Department of Physics, Technical University of Denmark, 2800 Kgs. Lyngby, Denmark}

\title[\texttt{achemso} demonstration]{Combining density functional theory with macroscopic QED for quantum light-matter interactions in 2D materials }

\usepackage{lineno}

\usepackage[nomarkers,nofiglist,figuresonly]{endfloat}
\begin{document}

\begin{abstract}
A quantitative and predictive theory of quantum light-matter interactions in ultra thin materials involves several fundamental challenges. Any realistic model must simultaneously account for the ultra-confined plasmonic modes and their quantization in the presence of losses, while describing the electronic states from first principles. Herein we develop such a framework by combining density functional theory (DFT) with macroscopic quantum electrodynamics, which we use to show Purcell enhancements reaching $10^7$ for intersubband transitions in few-layer transition metal dichalcogenides sandwiched between graphene and a perfect conductor. The general validity of our methodology allows us to put several common approximation paradigms to quantitative test, namely the dipole-approximation, the use of 1D quantum well model wave functions, and the Fermi's Golden rule. The analysis shows that the choice of wave functions is of particular importance. Our work lays the foundation for practical ab initio-based quantum treatments of light–matter interactions in realistic nanostructured materials.

\end{abstract}

\section{Introduction}

The development of van der Waals (vdW) heterostructures composed of atomically thin 2D layers has opened new opportunities for manipulating and enhancing light-matter interactions at the nanoscale. The very tight confinement of the electrons in these materials endows them with unique physical properties that can be altered and controlled by varying the number, type, and sequence of the atomic layers\cite{geim2013van,novoselov20162d}. For example, the experiment, which served as motivation for the current theoretical work, demonstrated thickness-tunable infrared emission via intersubband transitions in vdW quantum wells (QWs) comprised of few-layer transition metal dichalcogenide (TMD) structures\cite{schmidt2018nano}. At the same time, vdW heterostructures can support light-matter hybrid modes such as plasmon-, phonon-, and exciton-polaritons\cite{basov2016polaritons,jablan2009plasmonics} with ultra-confined electromagnetic fields in the direction normal to the 2D plane. The coupling between such polaritonic light modes and electron systems in nearby layers of the heterostructure provides an interesting avenue for controlling the optical properties of the material or even creating new types of quantum states -- if sufficiently strong coupling can be realized. However, the theoretical description and, in particular, the practical computation, of such coupled quantum light-matter systems present significant challenges that call for new developments\cite{koppens2011graphene, rivera2016shrinking,kurman2020tunable}. \\

\begin{figure}
    \centering
    \includegraphics[width=0.9\textwidth]{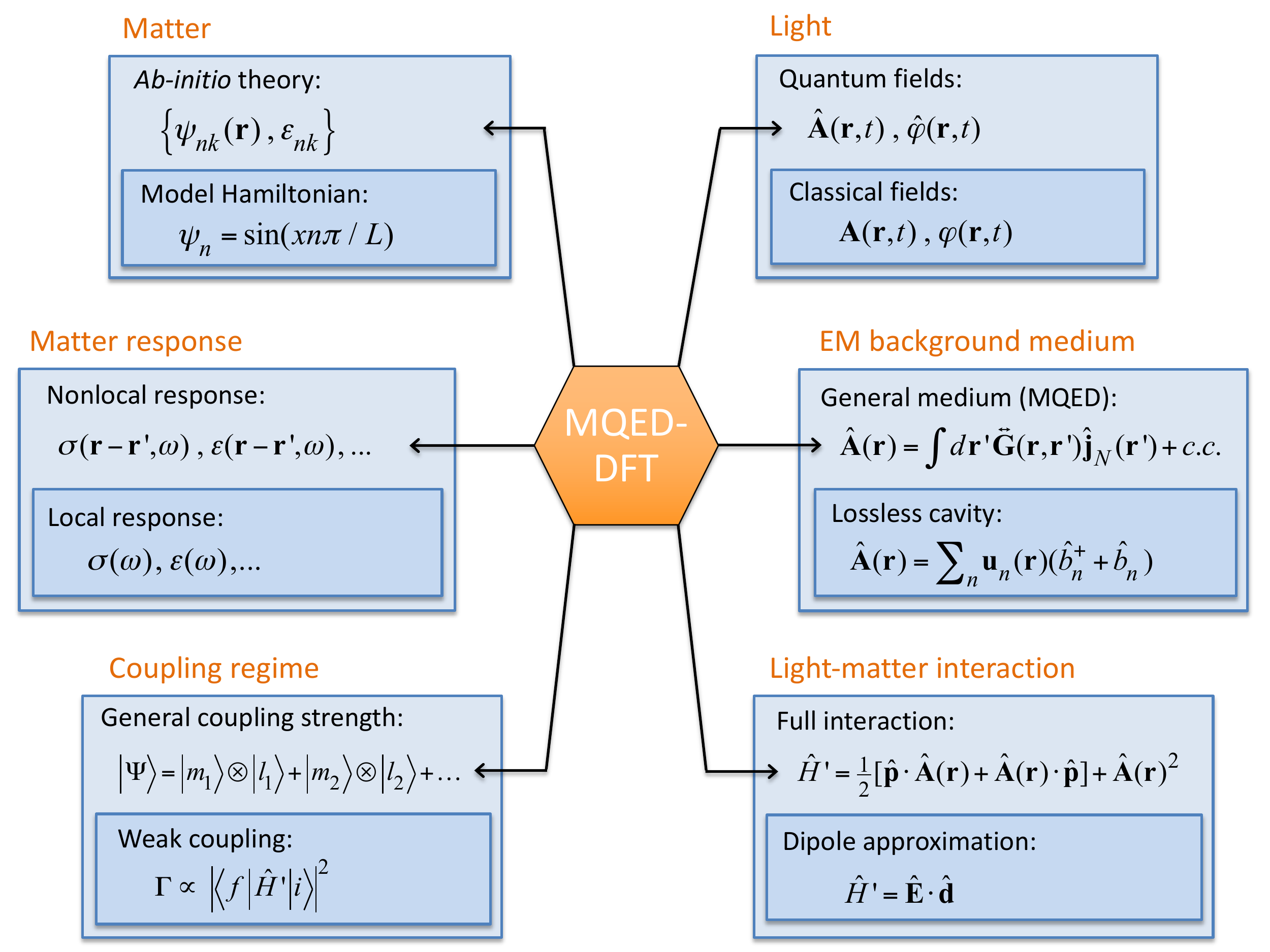}
    \caption{Schematic illustration of the key quantities (blue boxes) entering the description of light-matter interacting systems. For each quantity, the most general/accurate level of the theory is shown at the top while commonly used approximations are shown below. }
    \label{fig:overview}
\end{figure}

To set the stage for the following discussions, we briefly outline the most central notions and approximations relevant for the description of light-matter interactions in nanoscale materials, see Fig. \ref{fig:overview}.
First, the light-matter interaction is often described within the dipole approximation, which is valid whenever the electromagnetic (EM) field can be assumed to be constant over the extent of the matter system, or more precisely, the extent of the relevant electronic wave functions. The dipole approximation works well in many cases but becomes problematic for confined EM modes, such as the plasmon polaritons considered in the present work\cite{andersen2011strongly,takase2013selection}.  In such cases, the detailed shape of the wave function (and the light field) must be taken into account when evaluating the light-matter interaction. The question of how detailed the description of the electron wave function must be remains largely unexplored, and obviously depends on the specifics of the considered system and the targeted accuracy. For the TMD/graphene heterostructures studied in this work, we find that the difference in the spontaneous emission rates obtained with model wave functions and full ab-initio wave functions, can differ by an order of magnitude and show clear qualitative differences as well.  

When cavities or plasmonic structures are used to select and enhance distinct light modes, it is possible to reach the strong coupling regime where the pure light and matter states become entangled. For this to happen, the light-matter coupling strength must exceed the damping rate of the light mode; or equivalently, the radiative lifetime of the matter state should be shorter than the lifetime of the light mode, such that emission-reabsorption processes can occur. We note in passing that the lifetime of the light mode is determined by the optical losses occurring in the relevant materials (see below). Under strong coupling conditions it is necessary to diagonalize the light-matter Hamiltonian, or alternatively evolve the state dynamics, using non-perturbative methods (in practice within a truncated Hilbert space).\cite{latini2019cavity,torma2014strong,reithmaier2004strong} However, in many cases of interest the coupling will be sufficiently weak that first-order perturbation theory, i.e. Fermi's Golden rule, applies. The combination of the dipole approximation and Fermi's Golden rule yields an electronic transition rate given by the product of the dipole strength and the local photonic density of states (DOS). In this case, owing to the equivalence of the local photon DOS and the EM self-field of a point dipole, the spontaneous emission can be perceived as a semi-classical effect that depends on the electron wave functions only through the effective dipole strength. On the other hand, when the EM field varies on the scale of the electron wave function, the dipole-approximation breaks down\cite{andersen2011strongly}, selection rules change\cite{takase2013selection, rivera2016shrinking} and the local semi-classical picture must be replaced by a full quantum treatment that explicitly incorporates the nonlocal nature of light-matter interactions.

The conventional approach to the quantization of the EM field is based on an expansion into orthogonal modes. However, in the presence of optical losses, i.e. when the conductivity of the EM background medium is finite in the relevant frequency range, the expansion into orthogonal modes is ill-defined and the quantization must be performed using alternative approaches such as macroscopic quantum electrodynamics (MQED)\cite{Scheel2008MQED} or expansion in quasi-normal modes\cite{van2012spontaneous}. Finally, it is often necessary to invoke the optical response functions of the involved materials. This is for example the case for the MQED formalism to be discussed below. While spatially local, i.e. $q$-independent, response functions are often employed, and indeed work well in extended systems, this approximation becomes questionable in nanoscale structures such as 2D materials\cite{raza2015nonlocal,andersen2015dielectric}. 

As will be shown, all of the common approximations used to describe quantum light-matter systems (see Fig. 1) can be violated in vdW quantum well devices. Consequently, these systems are ideal as testbeds for the testing and development of theoretical methods in this exciting field.

In recent years, there has been significant developments toward a full ab-initio description of coupled light-matter systems based on the fundamental Hamiltonian of non-relativistic QED, also known as the Pauli-Fierz Hamiltonian\cite{Ruggenthaler2018review}. In particular, the quantum-electrodynamical density functional theory (QEDFT), which is a density functional type reformulation of the Pauli-Fierz Hamiltonian\cite{flick2015kohn}, has been introduced as a generalization of time-dependent density functional theory to cases where the quantum nature of the light is important.\cite{Ruggenthaler2018review, ruggenthaler2014quantum} The fundamental nature of the QEDFT formalism has, for example, enabled studies of atomic emitters in cavities where the back-action of the EM field leads to "dressing" of the electronic states.\cite{flick2017SC,Wang2020WeaktoStrongLC} Although general, the QEDFT approach still has its limitations. An important one for making quantitative predictions of light-matter interactions in real materials, is the assumption of a lossless\cite{Tokatly2013PRL} or low-loss\cite{neuman2018coupling} EM background medium. Although this problem can in principle be alleviated by including the lossy environment as part of the active matter system\cite{Ruggenthaler2018review} it is unlikely that this would be practical for complex nano-optical setups, such as those involving vdW-heterostructures.\cite{koppens2011graphene, geim2013van} We further note that most applications of QEDFT have been restricted to the dipole approximation, although steps towards a more general description have recently been taken.\cite{neuman2018coupling,jestadt2019light} Finally, most QEDFT studies have been restricted to finite systems, such as atoms and small molecules, while the application to extended systems has been limited.

In this paper, we introduce an approach based on the Wigner-Weisskopf model to describe quantum light-matter interactions in complex 2D material setups under quite general conditions. Our methodology, combines MQED for quantizing the optical modes of an arbitrary optical environment with the single particle Kohn-Sham wave functions and energies from DFT for the electrons. This combination enables a quantitative description of the physics in a regime where quantized light modes, confined to a length scale of the electron wave function, interact strongly with the electrons while being subject to losses in the EM background medium. We use the method to calculate the radiative transition rates of a multilayer TMD quantum well sandwiched between graphene and a metal surface, see Fig. 2a. By varying the type of TMD material, the number of layers, and the graphene doping level, we determine the maximal Purcell enhancement achievable in such a setup to be $10^7$, corresponding to intersubband transition rates of 4.5 THz with high efficiency of emission into the propagating plasmons. For the thinnest 2-layer stacks we observe somewhat lower Purcell factors of around $10^6$, but higher absolute rates approaching 30 THz. These radiative lifetimes are at least an order of magnitude lower than the intrinsic (electron-phonon limited) lifetimes of electrons and holes in TMDs at low temperatures\cite{kaasbjerg2013acoustic,hinsche2017spin,schmidt2018nano}. Consequently, the coupling to the graphene plasmons should indeed be the dominant mechanism governing the dynamics of the intersubband transitions at low temperatures.
We compare our DFT-MQED results with common approximation schemes such as the dipole approximation and the use of 1D quantum well wave function models. We find quantitative differences in the obtained rates of up to an order of magnitude. In particular, we find that the rates are highly sensitive to the detailed spatial shape of the wave function, emphasizing the need for a proper description of the electronic states. Lastly, we explore cases where the first-order QED breaks down and higher-order processes must be considered to obtain a correct description of the Purcell effect.

\begin{figure}
    \centering
    \includegraphics[width=\textwidth]{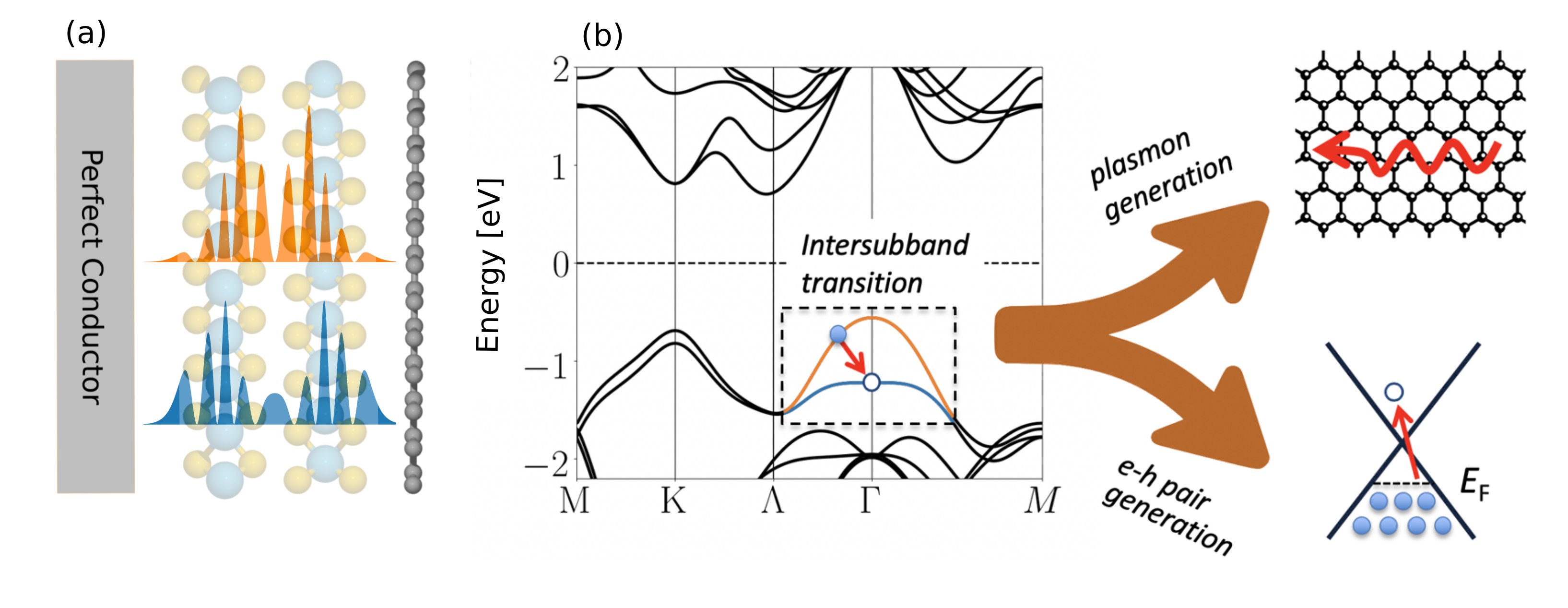}
    \caption{\textbf{Combining DFT and MQED in van der Waals materials.} (a) Illustration of a 2-layer TMD quantum well sandwiched between a perfect conductor and a doped graphene sheet. The $z-$profile of two subband wave functions obtained from DFT are also shown. (b) The DFT band structure of bilayer WSe$_2$ with the two subbands at the $\Gamma$-point indicated in orange and blue. An electron from the filled upper subband can combine with a hole at the top of the lower subband by emitting either an electron-hole pair (loss) or a propagating acoustic plasmon-polariton.}
    \label{fig:setup}
\end{figure}

\section{Results}
\subsection{The MQED-DFT framework}
The framework described in the following section is motivated by the new regimes of light-matter interactions that are now accessible in nano-materials such as vdW heterostructures. Although the presented framework is general, we shall focus on the case of 2D layered geometries and apply the formalism to a TMD quantum well (QW) sandwiched between a conductor and a graphene sheet, see Fig. 2(a). Such heterostructures feature rapid emission from the QW into acoustic graphene plasmons\cite{iranzo2018probing,iranzo2018science,lundeberg2017tuning} that require a beyond-dipole MQED framework with ab-initio QW wave functions. We note that our work is the first to combine MQED with DFT. Moreover, the use of the Wigner-Weisskopf model enable us to analyze cases within the strong-coupling regime. A similar use of the Wigner-Weisskopf model can be found in \cite{Vogel2006,neuman2018coupling}, which also go beyond the Fermi’s golden rule. We present here the main steps of our framework, while the details are described in Supplementary note 1.

We represent the light-matter interaction Hamiltonian in the Weyl gauge (which is defined by the requirement of a vanishing scalar potential),

\begin{equation}\label{eq:Hint}
\hat{H}_\mathrm{int} = \frac{e}{m}\mathbf{\hat{A}}\cdot\mathbf{\hat{p}} - \frac{ie\hbar}{2m}\boldsymbol{\nabla}\cdot\mathbf{\hat{A}}\, ,
\end{equation}

\noindent where $e$ is the electron charge, $m$ is the electron mass, and $\mathbf{\hat{p}}$ represents the electron momentum operator. In writing Eq. (\ref{eq:Hint}) we neglected the diamagnetic $\mathbf{\hat{A}}^2$ term. The justification of this assumption and its implications are discussed in the "Limitations of MQED-DFT" subsection below. The Weyl gauge is an incomplete gauge and should therefore be accompanied by an additional condition to completely fix the gauge to avoid unphysical excitations arising in the quantization scheme\cite{hatfield2018quantum}. We mention in passing that this condition is met because $\boldsymbol \nabla \cdot (\tensor{\epsilon}\cdot \mathbf{A}) = 0$ inside of the anisotropic TMD layer, the justification of which is discussed later. We further note that this means that in an isotropic dielectric material, our gauge reduces to the normal Coulomb gauge ($\boldsymbol \nabla \cdot \boldsymbol A = 0$) common in quantum optics. In such cases, the $\boldsymbol{\nabla}\cdot\mathbf{\hat{A}}$ term disappears from Eq. \ref{eq:Hint}, leaving only the $\mathbf{\hat{A}}\cdot\mathbf{\hat{p}}$ term. The $\boldsymbol{\nabla}\cdot\mathbf{\hat{A}}$ term can thus be thought of as an anisotropic correction and we have generally found it to be small relative to the $\mathbf{\hat{A}}\cdot\mathbf{\hat{p}}$ term.

Within MQED in the Weyl gauge, the vector potential is expressed as\cite{glauber1991quantum,raabe2007unified,novotny2012principles}, 

\begin{align}\label{eq:vector_potential_local}
    & \hat A_i(\mathbf{r}) =  \sqrt{\frac{\hbar}{\pi \epsilon_0}}\int d\omega \frac{\omega}{c^2}\int d^3 s \,  \tensor{G}_{il}(\mathbf{r}, \mathbf{s}, \omega)\left[\sqrt{\textrm{Im}\, \epsilon(\mathbf{s}, \omega)}\right]_{lj}\hat{f}_j(\mathbf{s}, \omega) + \textrm{h.c.} \, ,
\end{align}

\noindent where $\hbar$ is the reduced Planck's constant, $\epsilon_0$ is the vacuum permittivity, $c$ is the speed of light, and the subscripts obey Einstein's summation convention. The quantization scheme is build around the Langevin noise current annihilation and creation operators, $\hat{J}(\mathbf{s},\omega)=\left[\sqrt{\textrm{Im}\, \epsilon(\mathbf{s}, \omega)}\right]_{lj}\hat{f}_j(\mathbf{s}, \omega)$ (and similar for $\hat{J}^{\dagger}$), with  $\epsilon_{lj}(\mathbf{s}, \omega)$ being the relative permittivity. The classical Dyadic Green's function (DGF), $\boldsymbol{\tensor{G}}(\mathbf{r}, \mathbf{s}, \omega)$, which satisfies  
\begin{align}
    \left[\boldsymbol{\nabla}\times \boldsymbol{\nabla}\times - \left(\frac{\omega^2}{c^2}\right) \boldsymbol{\tensor{\epsilon}}(\mathbf{r}, \omega) \right] \boldsymbol{\tensor{G}}(\mathbf{r}, \mathbf{s}, \omega) = \boldsymbol{\tensor{I}} \delta(\mathbf{r} -\mathbf{s}),
\end{align}

\noindent gives the electric field at position $\mathbf r$ generated by a point dipole of frequency $\omega$ at position $\mathbf s$ in the presence of an EM background medium with permitivity $\tensor{\epsilon}$. In our case, the EM background permittivity is defined by the perfect conductor, the graphene sheet, and the TMD without contributions from the subband transitions (these define the active matter system, see below, and should be excluded from $\tensor{\epsilon}$ to avoid double counting). In general, determining the DGF is a nontrivial task, which is often accomplished using numerical methods \cite{lalanne2019quasinormal}. However, in our case, the layered geometry of the vdW-heterostructure and the accompanying in-plane translation symmetry makes it possible to obtain a closed analytical solution for the DGF (see the Methods section).

The active matter system is defined by the electrons in the valence subbands near the $\Gamma$-point of the TMD quantum well, see Fig. \ref{fig:setup}(b). We determine the electronic structure of the subbands of the isolated TMD structure by means of DFT calculations. DFT allows us to access the band structure and the single particle wave functions of the subbands\cite{dreizler}, which in turn are used to evaluate the coupling matrix elements of the light-matter interaction Hamiltonian as described below. The details of the DFT calculation are provided in the Methods section.

Due to the in-plane translational invariance of the system, the electronic states of the subbands can be labelled by a subband index and a wave vector. In the following we consider the process where an electron from the upper subband (orange band in Fig. 2(b)) recombines with a hole at the top of the lower subband (blue) with the EM field initially in its ground state, i.e. a spontaneous emission process. The rate of this transition can thus be seen as the inverse lifetime of a hole in the lower subband due to the coupling to the EM field.  Denoting the ground state of the TMD by $|\Psi_0\rangle$, the initial state representing a hole at the top of the lower subband ($l$) is written $|i\rangle=\hat c_{l,\mathbf 0}|\Psi_0\rangle$, while the final state with a hole in the upper subband ($u$) at momentum $\mathbf q$ is given by $|f_\mathbf q\rangle = \hat c_{u,\mathbf q}|\Psi_0\rangle$. 

The time-dependent combined state of the quantum well and EM field takes the form

\begin{align}\label{eq:timedp}
|\Psi(t)\rangle = C_{i}(t)e^{-i\omega_it}|i\rangle\otimes |0\rangle+ \sum_{\mathbf q}\sum_{\nu_\mathbf q} C_{f,\nu \mathbf q}(t)e^{-i(\omega_{\nu\mathbf{q}} + \omega_{f \mathbf q})t}|f_\mathbf q\rangle\otimes|\nu_\mathbf q\rangle \, ,
\end{align}

\noindent where $|\nu_\mathbf q\rangle$ labels a single photon state of the EM field, which due to the in-plane translational invariance of the layered structure considered in this work is described by a continuum of energies and in-plane momenta. We note that with the form of the many-body state in Eq. (\ref{eq:timedp}) we are limited to one-photon final states and thus we neglect multi-photon processes. Using the Wigner-Weisskopf approach, it can be shown that the time evolution of the initial state coefficient satisfies the equation of motion

\begin{align}\label{eq:kernel_Ce}
    \dot{C}_{i}(t) = -\int_0^t dt' \, \int d\omega \int d\boldsymbol{q} K(\omega, \boldsymbol{q})e^{-i(\omega-\omega_{if \boldsymbol{q}})(t-t')}C_{i}(t'),
\end{align}

\noindent where $\hbar \omega_{if \mathbf q}=\varepsilon_{f\mathbf q}-\varepsilon_i$ is the electronic transition energy and the coupling kernel is given by

\begin{align}\label{eq:K}
    K(\omega, \boldsymbol{q}) = -\frac{1}{4\pi^2}\frac{e^2}{4m^2}\sum_{\nu_\mathbf q} \left|\langle f_\mathbf {q} \otimes \nu_\mathbf{q} |\hat H_{\mathrm{int}}|i\otimes  0 \rangle\right|^2\delta(\omega-\omega_{\nu \mathbf q}). 
\end{align}

The Wigner-Weisskopf approach defined by Eqs. (\ref{eq:kernel_Ce}) and (\ref{eq:K}) describes the decay of the excited state into the EM environment and includes all optical one-photon decoherence channels. We note that via proper generalization other loss and decoherence sources, such as electron-phonon coupling, not considered in the current work, could in principle be accounted for by appropriate generalization of the many-body state and the coupling kernel. To account for pure dephasing a density matrix representation of the state is required and the Hilbert space would have to be extended to include the sets of states, $\left\{|i\rangle\otimes|\nu_\mathbf q\rangle\right\}$ and $\left\{|f_\mathbf q\rangle\otimes|0\rangle\right\}$.

\begin{figure}
    \centering
    \includegraphics[width=\textwidth]{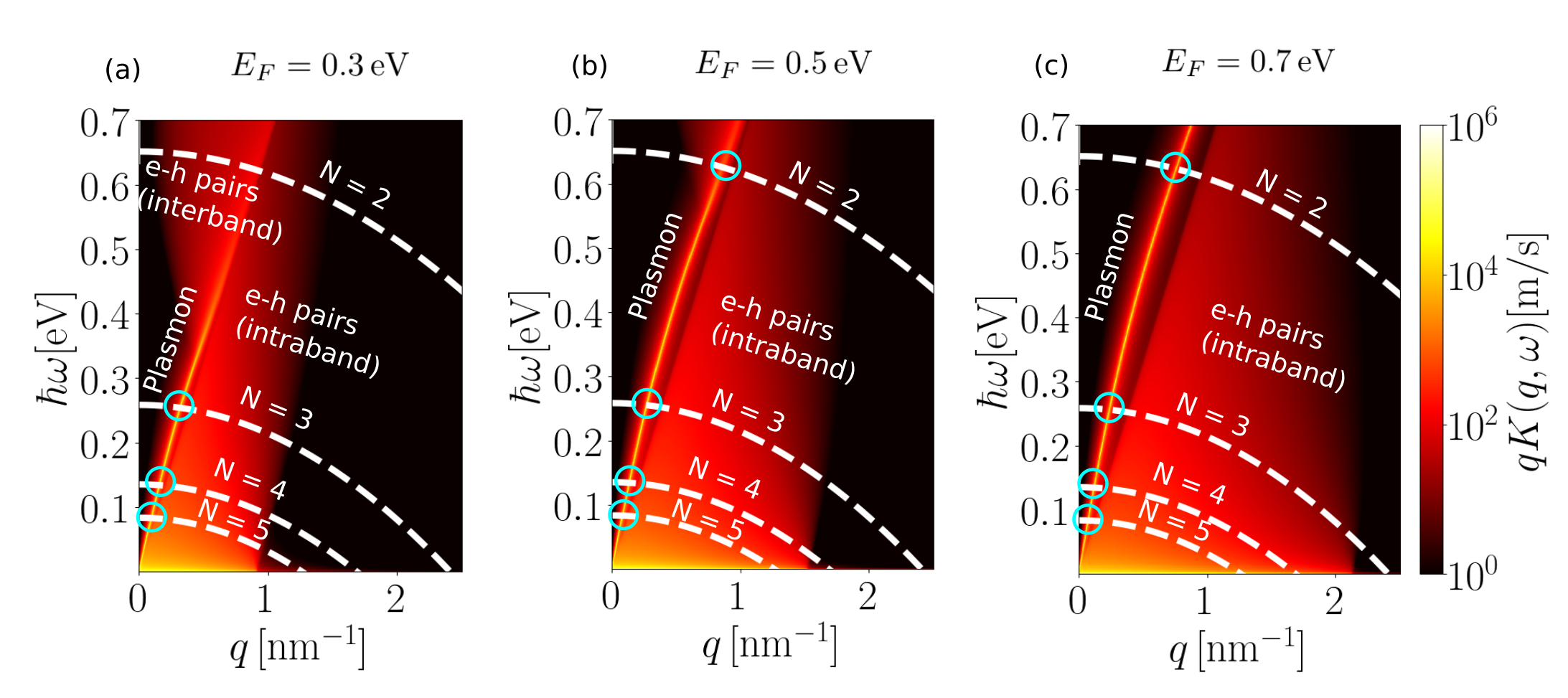}
    \caption{\textbf{Overlapping the electronic bandstructure and the plasmon dispersion to calculate their light-matter interactions.} The coupling kernel for the transition between of holes between the lower and upper subband in a two-layer $\mathrm{WSe}_2$ quantum well placed between a perfect conductor and a doped graphene sheet is shown for three different values of the graphene Fermi level, (a) 0.3 eV, (b) 0.5 eV and (c) 0.7 eV. The kernel describes the coupling into all optical excitations, including the plasmon and the electron-hole continuum as described by graphene's full nonlocal conductivity\cite{gonccalves2016}. The dashed lines represent the non-vertical emission energies, $\hbar\omega_{if\boldsymbol{q}}$, derived from the DFT band structure for $\mathrm{WSe}_2$ quantum wells of $N$ = 2, 3, 4 and 5 layers and the light blue circles denote the intersection between the dispersion of the intersubband transition energy and the graphene plasmon.}
    \label{fig:Kernel_and_wfs_illustration}
\end{figure}

The kernel, $K(\omega, \mathbf q)$, in Eq. (\ref{eq:K}) describes the coupling strength between the electronic subband transitions and the optical excitations with frequency $\omega$ and in-plane momentum $\mathbf{q}$. To properly account for the nonlocality of the graphene, we employ the full nonlocal conductivity of the graphene\cite{gonccalves2016} as described in Methods. Importantly, we model the graphene via a nonlocal surface conductivity, neglecting the thickness of the graphene sheet. We note that treating the graphene as a 3D material with a finite thickness (with a bulk conductivity instead of a surface conductivity) should not affect the graphene plasmonic properties and thus not affect the results. The graphene conductivity depends strongly on the graphene Fermi level. The kernel therefore has to be calculated independently for each value of the graphene Fermi level. Exploiting the in-plane isotropy of the subband band structures, in Fig. \ref{fig:Kernel_and_wfs_illustration} we show a color map of the coupling kernel integrated over the polar angle, $qK(\omega,q)$,  for a 2-layer WSe$_2$ quantum well at three different values of the graphene Fermi level, alongside the transition energies, $\omega_{if\boldsymbol q}$, for the different numbers of layers. The acoustic graphene plasmon and the dissipative electron-hole (e-h) continuum are clearly visible. 

From Eq. (\ref{eq:K}) it follows that the magnitude of the kernel at a given $(\omega, \mathbf q)$ depends on two factors: the spatial overlap of the subband wave functions and the electric field of the optical mode (the matrix element), and the number of available optical modes at $(\omega, \mathbf q)$ (the photonic DOS). The graphene plasmons affect the kernel via large matrix elements originating from their strong EM field. On the other hand, the e-h excitations in graphene individually produce weak EM fields but due to their large number they still can have a significant influence on the kernel, especially at small frequencies compared to the Fermi-level.

There are several aspects of the formalism worth highlighting. First, we emphasize that the normal modes of the EM field, $|\nu_\mathbf q\rangle$, are used only symbolically in the above equations. In fact, as mentioned previously, the normal modes are not well defined in the presence of lossy media like the graphene sheet. Instead, one has to resort to so-called quasi-normal mode formalisms, of which MQED is one example, but others exist\cite{ren2020near}. In practice, the MQED formalism allows us to obtain the kernel in terms of the DGF. Focusing on the dominant $\mathbf{\hat{A}}\cdot\mathbf{\hat{p}}$ term from the interaction Hamiltonian for simplicity, it can be shown that (see Supplementary note 1),
\begin{align}\label{eq:matrix}
     \sum_{\nu_\mathbf q}  \left|\langle f_\mathbf {q} \otimes \nu_\mathbf{q} |\hat H_{\mathrm{int}}|i\otimes  0 \rangle\right|^2   \delta(\omega-\omega_{\nu \mathbf q}) = 
    \int \int d\mathbf r d \mathbf r'\, p^*_k(\mathbf r) \mathrm{Im}\, \tensor{G}_{kl}(\mathbf r,\mathbf r',\omega)p_l(\mathbf r')\, ,
\end{align}

\noindent where we have defined the transition current $p_l(\boldsymbol r) = [\psi^*_{f\mathbf q}(\mathbf r) \hat{p}_l\psi_i(\mathbf r) ]$. By virtue of the DGF, an expansion in normal modes becomes superfluous, and this resolves the problem of ill-defined normal modes. However, the replacement of the normal modes by the DGF is not only a technicality; it also leads to a somewhat different picture of the quantum light-matter interactions itself. The left hand side of Eq. (\ref{eq:matrix}) describes coupling of electronic states $|i\rangle$ and $|f_\mathbf q\rangle$ via the manifold of degenerate EM modes at frequency $\omega$. In this picture, the EM field driving the electronic transition is simply the field of the EM vacuum, i.e., $\hat{\mathbf{A}}|0_{\mathrm{EM}}\rangle$. In particular, it is independent of the state of the electrons. A quite different picture is suggested by the right-hand side. Recalling that $i\tensor{\mathbf G}(\mathbf r,\mathbf r',\omega)$ is the electric field at $\mathbf r$ created by a point current source at $\mathbf r'$, we can interpret the double integral as the EM self-energy of the current distribution $\langle f_{\mathbf q}|\hat{\mathbf{p}} (\mathbf r')| i\rangle$ (we must use the imaginary part of the DGF to obtain the real part of the energy). We can further notice that $\langle f_{\mathbf q}|\hat{\mathbf{p}} (\mathbf r')| i\rangle$ is a measure of the quantum fluctuations of $\hat{\mathbf{p}} (\mathbf r)$ in the initial and final states (the matrix element vanishes if one of the states is an eigenstate of $\hat{\mathbf{p}} (\mathbf r')$).  Therefore, in this picture, the electronic transition is driven by an EM field whose source is the current density of the transition itself.

\subsection{Limitations of MQED-DFT}\label{sec:limit}
Although the presented framework is quite general it still has its limitations. Most importantly, the framework is limited to the weak and strong coupling regimes, but not suited for treating cases of ultra-strong coupling. The first reason is that the framework is built around the Wigner-Weisskopf model which means that we are restricted to a one-photon Hilbert space and that we neglect the counter-rotating terms. Both of these assumptions are expected to be violated in the ultra-strong coupling regime\cite{kockum2019ultrastrong}.
Secondly, the fact that the wave functions and energies of the active matter system (the TMD subbands in the present case) are calculated independently of the vacuum fluctuations means that the states of the system might be misrepresented in situations where there is significant dressing of the electronic states by the vacuum field. In such cases, one should resort to a fully self-consistent method such as e.g., QEDFT. We stress, however, that such situations are only expected to arise in the case of ultra-strong coupling, which is \textbf{very} difficult to reach without external pumping of the EM field. Within the typical regimes of weak to strong coupling, in particular this includes systems in equilibrium, i.e., non-driven systems, we thus expect the MQED-DFT method to be valid and provide an efficient alternative to more full-fledged ab-initio treatments. 

To verify that the dressing of the subband states considered in this work is indeed small, we calculated the Lamb-shifts of the electronic energies and found that they never exceed 0.4 \% of the transition energy, (see Supplementary note 5). We thus conclude that the dressing of the electronic states by the EM vaccum field is neglible, and that a non-self consistent treatment of the light-matter interaction is justified for the systems studied in the present work.

Next we discuss the gauge fixing condition $\nabla\cdot\left(\tensor{\epsilon}\cdot \mathbf{A}\right) = 0$. This condition is met since the MQED vector potential is defined in the absence of the intersubband transition and we assume that no other longitudinal sources exist inside of the TMD layer. The condition also follows directly from the form of the DGF presented in the methods section. This extra condition completely fixes the gauge and leaves us with a generalized version of the Coulomb gauge common in quantum optics. We note that the assumption that no other longitudinal sources exist is justified inside the TMD. At the TMD/graphene and TMD/conductor interfaces this condition is violated and the field develops a longitudinal component. However, the subband wave functions are well confined within the TMD region and have only negligible weights at the interfaces.

Next, we consider limitations arising from the chosen form of the interaction Hamiltonian used in this work. As mentioned in the above, in writing Eq. \ref{eq:Hint} we neglected the diamagnetic $\mathbf{\hat{A}}^2$ term. The diamagnetic current density, the $\mathbf{\hat{A}}^2$ term, has been neglected since it only describes two-photon emission/absorption and emission-reabsorption processes. The two-photon processes are not captured by our one-photon Hilbert space and neglecting them is consistent with the use of the Wigner-Weisskopf model. The emission-reabsorption processes leads to an energy renormalization of the states similar to the Lamb shift. We have calculated this shift using first order perturbation theory for a few of the considered structures and found it to be below 0.4\% of the undressed transition energies and thus negligible, see Supplementary note 4 for details on the calculation. 
While we believe that for the situation and method considered in this work the values calculated above give a good indication of the importance of the diamagnetic term, such an approach is only valid in a perturbative or finite level approach such as the restricted Hilbert space considered in this work\cite{schafer2020relevance}.
More generally, the $\mathbf{\hat{A}}^2$ term becomes dominant when reaching the ultra-strong coupling regime\cite{kockum2019ultrastrong} and any attempt at treating this regime would require an extension of the model to include the diamagnetic term. We further note that in any self-consistent treatment of both light and matter, e.g. QEDFT, the diamagnetic term is essential to ensure the stability of the coupled system even at weak coupling strengths\cite{schafer2020relevance} and any such approach must therefore naturally include the $\mathbf{\hat{A}}^2$ term.

Another limitation of our approach concerns the description of the EM field distribution via the DGF, which is based on a continuum model of the underlying materials, e.g. a (nonlocal) sheet conductivity for graphene and infinitely sharp interfaces between the materials. As a result, while the modes of the field resolve the mesoscopic structure of the vdW-heterostructure, the local fields due to the atomic structure are not included in the modes of the EM field.

Finally, we note that the assumption of a perfectly conducting metal substrate is an approximation. In practice, a finite conductivity would allow the field to penetrate into the metal and cause additional optical losses. While we expect these effects to be small in the spectral region considered here\cite{iranzo2018science}, they could be significant in e.g., the optical regime and can be included in the formalism at the price of a more elaborate (numerical rather than analytical) determination of the DGF.

\subsection{Decay rates from MQED-DFT}
In this section, we use the MQED-DFT framework to analyse the coupling of intersubband transitions to propagating graphene plasmons in the vdW quantum well devices depicted in Fig. \ref{fig:setup}. Throughout the discussion we shall distinguish the \emph{radiative} decay rate due to emission of propagating plasmons from the \emph{total} decay rate, which will also include the incoherent coupling to 
the e-h continuum in the graphene. The radiative rate is obtained by setting the coupling to the e-h continuum to zero (i.e. setting $K=0$ everywhere in the $(q,\omega)$-parameter space except in the region where the plasmon crosses the subband transition, see Fig \ref{fig:Kernel_and_wfs_illustration}). Although the electron is emitting to different modes (for example different propagation directions), the interaction is expressed in terms of a single decay rate that incorporates all the emission channels.

\begin{figure}
    \centering
    \includegraphics[width=\textwidth]{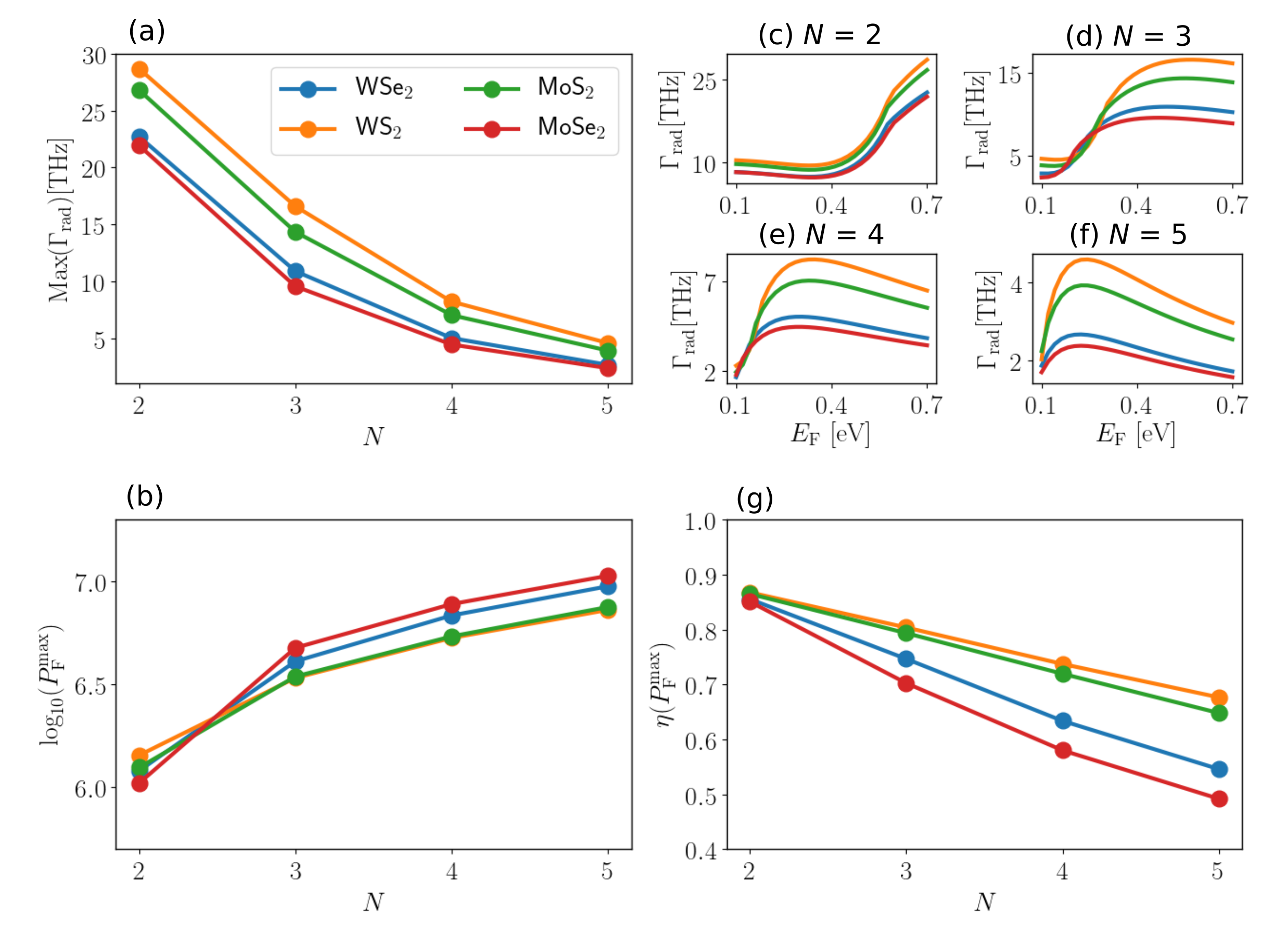}
    \caption{\textbf{The enhancement of transition rates as a function of number of layers and their composition.} (a) The maximum radiative rates obtainable by adjusting the graphene Fermi level, for the different TMD materials and number of layers in the quantum well ($N$). (b) The Purcell factors corresponding to the maximal radiative rates in (a). (c-f) Radiative emission rates as a function of graphene Fermi level for 2(c), 3(d), 4(e) and 5(f) layers. (g) Coupling efficiency, $\eta = \Gamma_{\mathrm{rad}}/\Gamma_{\mathrm{tot}}$, evaluated for $E_F$ giving the maximum radiative rate. All rates were calculated using the generalized Weisskopf-Wigner MQED-DFT framework.}
    \label{fig:Max_Radiative_rates}
\end{figure}

In Fig. \ref{fig:Max_Radiative_rates} we report the calculated radiative emission rates and Purcell factors for quantum wells of MoS$_2$, MoSe$_2$, WS$_2$, and WSe$_2$ as a function of the number of layers ($N=2,..,5$) and the graphene doping level ($E_F=0.1-0.7$ eV). 
In general, the four TMDs show very similar behaviour with the (small) differences mainly originating from variations in the subband energy gaps. Fig. \ref{fig:Max_Radiative_rates}(a,b) shows the maximum radiative emission rate and corresponding Purcell factor obtainable by tuning of $E_F$. The decay rates range from 2 THz to 28 THz, with the largest rates obtained for the thinnest structures. The corresponding lifetimes are at least an order of magnitude shorter than the intrinsic (electron-phonon limited) lifetimes of electrons and holes in TMDs at low temperatures\cite{kaasbjerg2013acoustic,hinsche2017spin,schmidt2018nano}. Consequently, the coupling to the graphene plasmons should indeed be the dominant mechanism governing the dynamics of the intersubband transitions, at least in the low temperatures regime. The higher rates obtained for thinner structures is somewhat counter-intuitive as the thinner structures have smaller transition dipole moments. However, this effect is dwarfed by the larger number of plasmons available at higher energies (the 2D acoustic plasmon DOS scales as $\sim \omega$) where we find the transition energies of the thinner structures, see Fig. \ref{fig:Kernel_and_wfs_illustration}. Another way to think of this is that the plasmon is more confined at the larger transition frequencies of the thinner structures leading to stronger field enhancement\cite{iranzo2018science}.

While the absolute decay rates decrease with the number of layers, $N$, the Purcell factors show the opposite trend. This goes against the common $Q/V$-scaling rule for the Purcell enhancement and can be explained as follows: The (vertical) intersubband transition energy scales as $\omega_{if}(N)\sim 1/N^2$. Within the point dipole approximation, the Purcell factor is essentially a measure of how much the photon DOS at the position of the emitter is enhanced relative to the free space value. While the 2D acoustic plasmon DOS scales as $\sim \omega$, the free space DOS scales as $\omega^2$. This means that even though the plasmonic DOS increases with energy, its magnitude relative to free space DOS actually decreases with energy. This argument therefore explains why the Purcell enhancement decreases with decreasing thickness. Going beyond the dipole approximation, another argument for the increase in Purcell factor with $N$, is the change in the matrix element due to the spatial variation of the plasmon along the QW. As we will show in Fig. 5, variations from the dipole approximation do occur in the thicker structures. However, since the deviations from the dipole approximation are relatively small, we deduce that the dominant effect leading to the larger Purcell factor in the thicker structures is the relatively larger plasmonic density of states at the intersubband transition energy as compared to its photonic counterpart.

As can be seen from Fig.  \ref{fig:Max_Radiative_rates}(c-f), the rates show a strong dependence on the graphene doping level. This behavior is not surprising as the magnitude and dispersion of the plasmon EM field increases with the density of free carriers. This explains the initial rise of $\Gamma_{\mathrm{rad}}$ with $E_F$ (except for the two-layer wells to which we return below). After the initial growth, the rates level-off and even decrease as $E_F$ is raised further. This happens because the dispersion curve of the graphene plasmon steepens as $E_F$ is increased. This steepening of the plasmon dispersion at larger $E_F$ results in a smaller transition momentum and thus smaller confinement of the plasmonic mode, leading to a reduction of the plasmonic density of states at the intersubband transition energy (blue circles in Fig. \ref{fig:Kernel_and_wfs_illustration}). Within the range of $E_F$ considered, the larger transition energies of the 2-layer stacks implies that the reduction of the plasmonic DOS due to the aforementioned dispersion steepening does not affect these thinnest structures, which explains why we do not observe a drop in the radiative rate for these systems.

Finally, we consider the radiative efficiency, $\eta = \Gamma_{\mathrm{rad}}/\Gamma_{\mathrm{tot}}$, evaluated for the graphene Fermi level giving the maximum Purcell enhancement, see in Fig. \ref{fig:Max_Radiative_rates}(g). The efficiencies range from 0.5-0.9 with the highest values attained for the thinner structures when the Fermi level is chosen to maximize the radiative emission rate. This trend can again be traced to the higher subband transition energies in the thinner structures, which implies that the dispersion curve $\omega_{if}(q)$ avoids the electron-hole continuum, see Fig. \ref{fig:Kernel_and_wfs_illustration}.
We note that since the metallic substrate is modelled as a perfect conductor there might be some additional losses arising from a real metallic substrate. However, we expect that these will be small in the infra-red region\cite{iranzo2018science} and indeed we have found the corrections from a real model of the metal to be small relative to the rates found here. Therefore, to sum up this chapter, using our formalism we can deduce that the most effective and the fastest light-emitting TMD device should be formed by only a 2-layer TMD.

\subsection{Assessment of common  approximations}
\begin{figure}
    \centering
    \includegraphics[width=\textwidth]{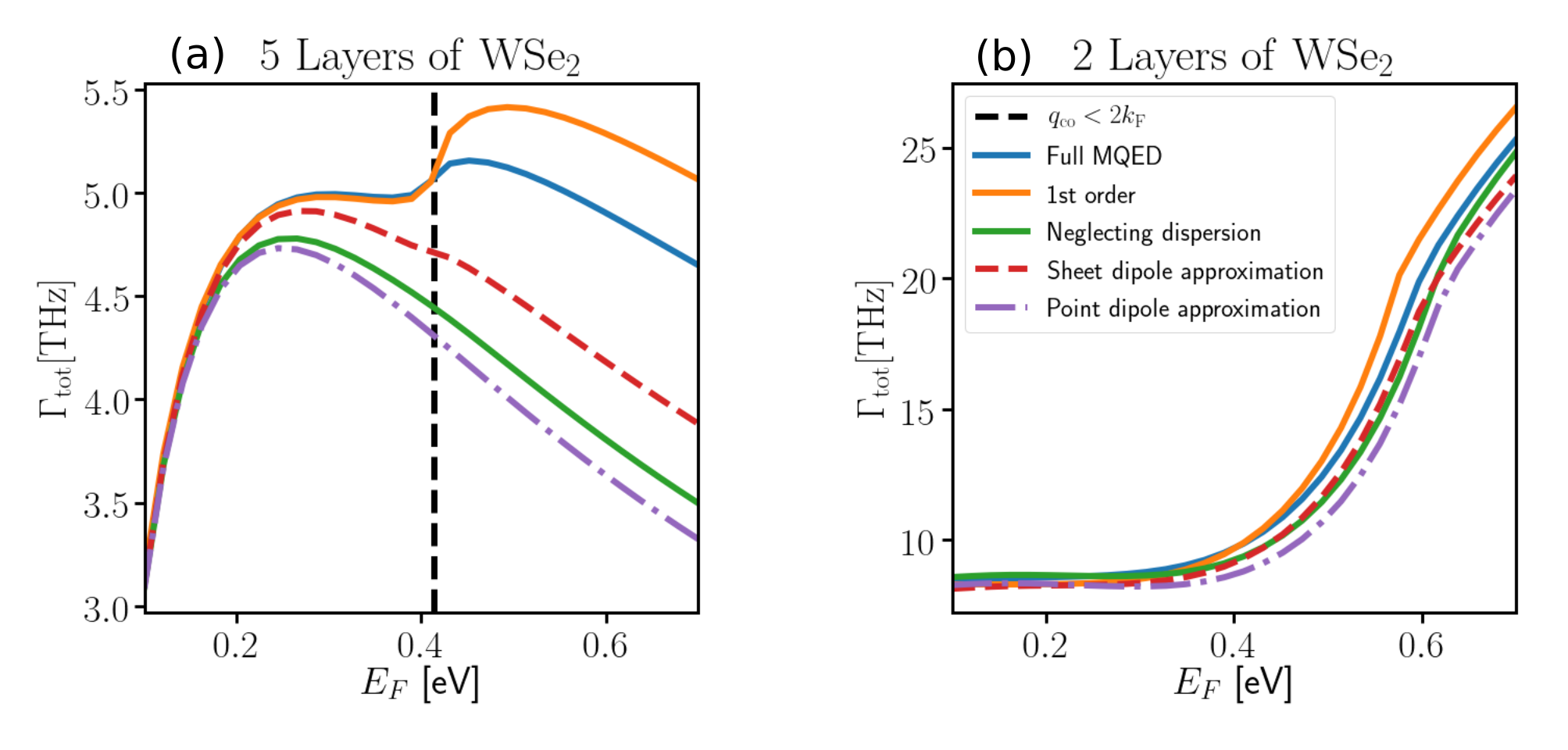}
\caption{\textbf{Comparison of the total rates calculated using the five different approximations.} The comparison is shown for 2 \textbf{(a)} and 5 \textbf{(b)} monolayers of $\mathrm{WSe}_2$: The full time dependent MQED framework (blue), 1st order MQED perturbation theory (orange), full MQED framework while neglecting the dispersion of the intersubband transition energy (green), approximating the TMD as a dipole sheet at $z_0 = d/2$ (red), and the point dipole approximation at $z_0 = d/2$ (purple). The dashed black line marks the graphene Fermi level at which horizontal transitions are opened up.}
\label{fig:Comparing_approximations}
\end{figure}

Having discussed the main results of the full MQED-DFT calculations, we now turn to a detailed analysis of the role played by the different elements of the method and a critical assessment of commonly used approximations. In this chapter we will focus on Fermi’s golden rule, the dipole approximation, and the possible choices for modeling the electronic wave functions when going beyond the dipole approximation. The discussion will be centered around Fig. \ref{fig:Comparing_approximations}, which shows a comparison of the total emission rate calculated with various approximations for a 5-layer (a) and 2-layer (b) WSe$_2$ quantum well.

\paragraph{First-order perturbation theory.}

We first consider the applicability of first-order perturbation theory, i.e. Fermi's Golden rule, for calculating emission rates. In terms of the coupling kernel, the expression for the spontaneous emission rate takes the form

\begin{align}\label{eq:FGR_expression}
    \Gamma = 2\pi\int_0^{2\pi}d\theta\int_0^{q_{\mathrm{hz}}(\theta)}dq\,q \, K(\omega_{ifq}, q),
\end{align}

\noindent which follows by making the Markov approximation in Eq. (\ref{eq:kernel_Ce}). In the momentum integral, the $q = 0$ limit corresponds to a vertical transition in the band structure while the limit $q = q_\mathrm{hz}$ denotes the wave vector at which the transition becomes horizontal. 
The argument of the kernel expresses the conservation of energy and in-plane momentum in all transitions. Consequently, within the 1st order picture, the emission rate is thus assumed to be the sum of the kernel values along the arc $(q,\omega_{ifq})$ (dashed white curves in Fig. \ref{fig:Kernel_and_wfs_illustration}). 

The blue curve in figure \ref{fig:Comparing_approximations}(a) shows the reference result of the full MQED-DFT method while the orange curve shows the 1st order result from Eq. (\ref{eq:FGR_expression}). As expected, the 1st order approximation works well for low $E_F$ where the plasmon is weak and the density of  e-h pairs is low. As $E_F$ is raised, the light-matter coupling becomes stronger, and beyond a certain threshold, marked by the vertical dashed black line, there is a clear deviation from the full-coupling reference curve. In this case, the back-action of the all the emitted light modes on the TMD electron actually prohibits a faster emission and as a result the 1st order perturbation theory overestimates the rate. The un-perturbed decay rate is found from the exponential decay fit of the excited state probability decay trend (See Supplementary figures 4 and 5). The breakdown of the 1st order perturbation theory occurs as a result of the coupling to low-energy horizontal e-h transitions across the graphene Fermi circle. Thus Fermi's golden rule is expected to fail when the horizontal QW transitions fall inside the e-h continuum, that is when $2k_F>q_\mathrm{hz}$. For the 5-layer QW in Fig. \ref{fig:Comparing_approximations}(a) this criterion occurs at approximately $E_F = 0.4 \mathrm{eV}$, while for the 2-layer structure in Fig \ref{fig:Comparing_approximations}(b) it occurs outside the plotted scale.

\paragraph{Dipole approximations.}
The next approximation we investigate is the local dipole approximation. Within the dipole approximation, the graphene assisted EM vacuum field driving the emission process is assumed to vary little over the extent of the intersubband wave functions, and Eq. (\ref{eq:matrix}) reduces to 
\begin{equation}\label{eq:dipole}
K(\omega, \boldsymbol{q}) \approx \hat{\mathbf p}_{if\mathbf q}^\dagger\cdot \mathrm{Im}\tensor{\mathbf G}(\mathbf r_0,\mathbf r_0,\omega)\cdot \hat{\mathbf p}_{if\mathbf q},
\end{equation}
where $\mathbf r_0$ is a point in the center of the QW and $\hat{\mathbf p}_{if\mathbf q}= \langle f_{\mathbf q}|\hat{\mathbf p}|i\rangle$ is the dipole matrix element evaluated from the DFT wave functions. Eq. (\ref{eq:dipole}) is the well known and widely used formula for spontaneous dipole emission\cite{novotny2012principles}. Below we test this strictly local approximation against the fully non-local expression Eq. (\ref{eq:matrix}). We consider two different types of dipole approximations, namely the conventional \emph{point-dipole} approximation and a \emph{sheet-dipole} approximation. To model a point-dipole emitter we neglect the $\mathbf q$-dependence of the dipole matrix elements and transition frequencies and replace them on the right-hand-side of Eq. (\ref{eq:dipole}) by their values at the vertical ($q=0$) transition. Unlike the point-dipole approximation, the sheet-dipole approximation takes the full $\mathbf q$-dependence of both dipole matrix elements (that is the QW wave functions) and transition frequencies into account, yet still evaluates the DGF at a single space point $\mathbf r_0$ in the center of the QW. (See Supplementary S1 for more details). Thus, the sheet-dipole approximation becomes exact for EM fields with $q$-dependent energy dispersion and spatial variation along the in-plane direction but not in the normal direction. 

\begin{figure}[!htb]
    \centering
    \includegraphics[width=\textwidth]{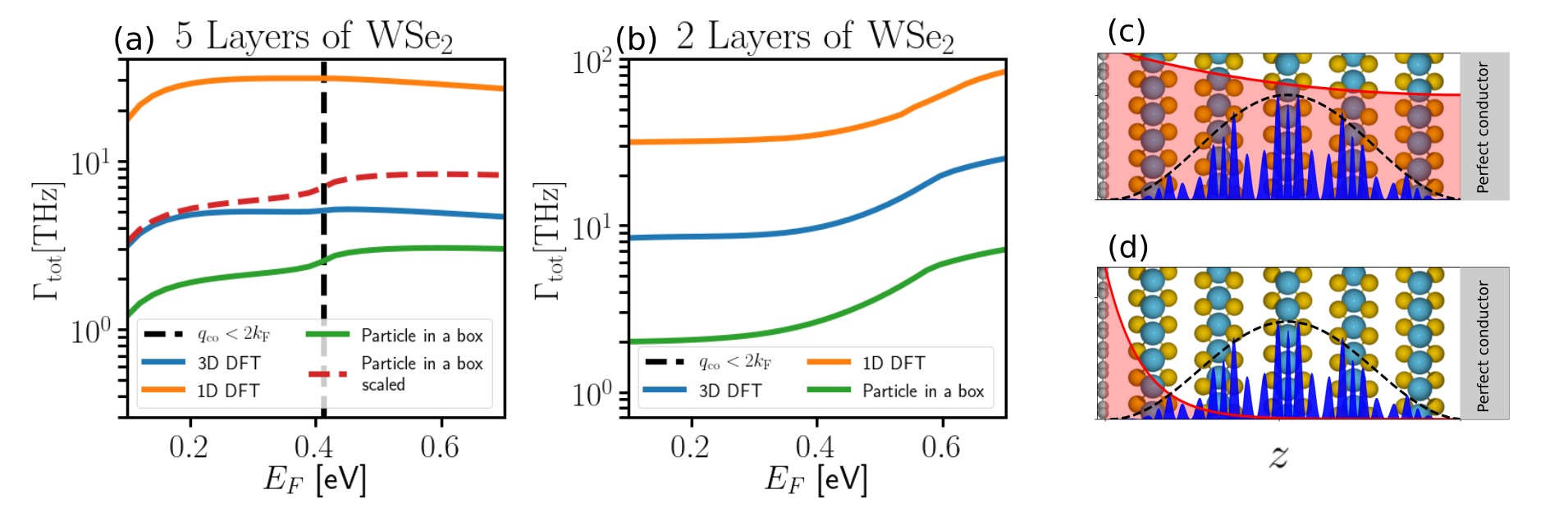}
\caption{\textbf{Comparison of the total emission rates calculated using three different wave function approximations: Particle in a box states, in-plane averaged DFT wave functions (1D DFT), and the full 3D DFT wave functions.} (a-b) Comparison of the total emission rates calculated with the different wave function approximations for a 2-layer and 5-layer WSe$_2$ quantum well, respectively. Additionally, the dashed red line in (a) shows the rates calculated using the particle in a box wave functions scaled by the square of the ratio between the transition dipole moments calculated with the full 3D DFT wave functions and the particle in a box wave functions. (c) and (d) The field enhancement profiles in red for the plasmon with in-plane wave vector $q = 0.2\mathrm{nm}^{-1}$(c) and $q= 1.21\mathrm{nm}^{-1}$(d). The latter corresponds to the wave vector for horizontal transitions, $q_{\mathrm{hz}}$, of the 5-layer $\mathrm{WSe}_2$ QW. The simple particle in a box wave function is shown by the dashed black line and the 1D DFT wave function is shown in blue. }
\label{fig:Comparing_wfs_approximations}
\end{figure}

The total decay rates obtained using the two dipole approximations are shown in Fig. \ref{fig:Comparing_approximations} by the purple (point-dipole) and red (sheet-dipole) curves, respectively. Focusing on the 5-layer structure, we observe that both dipole approximations work well for small $E_F$. In this regime, the decay is dominated by the coupling to the plasmons whose electric field is long range and varies little over the extent of the QW, thus fulfilling the assumption of the dipole approximation.  
However, for larger values of $E_F$, the point-dipole approximation deviates from the full-coupling reference and underestimates the emission rate significantly. This difference arises mainly due to the neglect of the $q$-dispersion of the transition energies, which implies that the coupling to the low-energy e-h pairs is missed. Indeed, we obtain very similar results by setting $\omega_{if}(\mathbf q)=\omega_{if}(\mathbf 0)$ in the full MQED calculation  (green curve). However, while the sheet dipole approximation indeed represents an improvement upon the point dipole approximation, it still falls short of the full MQED treatment. The $q$-dispersion of the transition energies alone is thus not enough to explain the shortcoming of the local treatment. Specifically, the coupling to the large $q$ excitations in the e-h continuum of graphene is also underestimated in the sheet dipole approximation, due to the rapid spatial variation of their associated electric fields across the QW width, see figure \ref{fig:Comparing_wfs_approximations} (d). In conclusion, any accurate description must therefore simultaneously account for the $q$-dispersion of the transition energies and describe the light-matter interaction beyond the dipole approximation.  We note that both the neglect of $q$-dispersion and the dipole approximation have little influence on the rate obtained for the 2-layer QW, because the transition energy is much higher and the emission is dominated by the plasmon, see Figure  \ref{fig:Kernel_and_wfs_illustration}.

\paragraph{Electronic wave functions.}
The final approximation that we investigate concerns the QW wave functions. As will be shown, different reasonable wave function choices can affect the rate by an order of magnitude. In Figure \ref{fig:Comparing_wfs_approximations}(a-b), we compare the total rates obtained using the full MQED framework with the full 3D DFT wave function (blue) to the results obtained using model wave functions of the form, $\psi(\mathbf r)=\chi(z)e^{i\mathbf k\cdot \mathbf r_{\parallel}}$, where $\chi(z)$ is either the in-plane averaged DFT wave function (orange) or a "particle in a box" approximation (green). To understand the role of the wave function, it is instructive to consider the explicit form of the light-matter coupling valid for our 2D geometry, 
\begin{equation}\label{eq:fullK}
    K(\boldsymbol{q}, \omega) \approx  C(\boldsymbol{q}, \omega)\times \left|\int_\mathrm{u.c.}\mathrm{d}\mathbf r_{\parallel}\int \mathrm{d}z\, u_{f\mathbf q}(\boldsymbol{r})\mathrm{cosh}(\sqrt{\frac{\epsilon_\parallel}{\epsilon_\bot}}qz)\partial_z u_i(\boldsymbol{r})\right | ^2.  
\end{equation}
In this expression $u$ denotes the (in-plane) periodic part of the QW wave functions and $C(\boldsymbol{q}, \omega)$ is independent of the wave functions (it contains the density and field strength of the graphene assisted EM vacuum modes, see the Supplementary note 1). 

The different wave function approximations produce two types of errors. The first type of error is a simple error in the effective dipole. This error occurs when the $q$ of all of the relevant excitations fulfill the condition $q \ll 1/d$ ($d$ is the QW width) such that the field is approximately constant over the width of the QW. This condition is fulfilled for the 2-layer well in Figure \ref{fig:Comparing_wfs_approximations}(a). In this case, the different wave functions produce different effective dipoles but with no  qualitative effect on the rate as function of $E_F$. As shown in the Supplementary note 2, the three curves in Figure \ref{fig:Comparing_wfs_approximations}(a) are to a very good approximation related by a simple rescaling by the square of their respective transition dipole moments.

The second type or error occurs when the wave functions are coupling to large $q$ excitations. In this case, the coupling becomes sensitive to the spatial shape of the wave functions. This is for example the case for the 5 layer structure, where the rate from the particle-in-a-box wave function deviates qualitatively from the DFT results for higher values of $E_F$. This can be clearly seen if we compare the rates calculated using the full 3D DFT wave functions to the rates calculated using the particle in a box states rescaled by the ratio of the the effective dipole moments, blue and dashed red lines in figure \ref{fig:Comparing_wfs_approximations}(a) respectively. Specifically, the qualitative deviation happens when $2k_F$ becomes comparable to $q_{\mathrm{hz}}$, as indicated by the vertical dashed line. At this point, strong quenching due to the coupling of the horizontal transition to the low energy electron hole pairs becomes possible. Importantly, however, the strong quenching is not observed for the DFT wave functions because they are localized further inside the TMD structure (further away from the interface with the graphene) as compared to the standing waves, and therefore their overlap with the vacuum field associated with the $2k_F$ e-h pairs is significantly suppressed, see Figure \ref{fig:Comparing_wfs_approximations}(c).

\section{Discussion}

In this paper, we have combined macroscopic quantum electrodynamics (MQED) with density functional theory (DFT) to obtain a computationally efficient framework for quantitatively describing quantum light-matter interactions in solid state systems. The active quantum matter is represented by its DFT band structure and wave functions while the electromagnetic background media are described by their appropriate macroscopic linear response functions, which can be spatially nonlocal and dissipative. While the Kohn-Sham energies and wave functions from DFT are expected to provide a good approximation to the true subband states, it should be mentioned that they are subject to the limitations of the specific exchange-correlation functional employed and the Kohn-Sham scheme in general. In the future it would thus be interesting to explore how the use of other wave function schemes, for example hybrid functionals or GW self-energies, would affect emission rates and other quantum light-matter properties. The approach taken in the present work is in many ways complementary to to the QEDFT formalism. While being less fundamental in nature than the QEDFT, our method avoids many of the approximations commonly made in the description of coupled light-matter systems, c.f. Figure \ref{fig:overview}, while it remains applicable to relatively large/complex systems. Instead of taking the ab-initio route directly, we take advantage of the fact that the quantization of the light degrees of freedom in interacting, potentially lossy light–matter systems can be achieved within the MQED formalism\cite{Scheel2008MQED,glauber1991quantum}. A powerful feature of the MQED framework is that it treats all light modes on an equal footing. Specifically, it achieves the simultaneous quantization of lossy- as well as undamped propagating modes.  While the coupling to propagating modes is almost always the main objective, in practice the optical losses are often the performance limiting factor. Since losses are present and often of significant practical importance in real materials, they are essential to include in a quantitative theory. 
Even within the MQED formalism, and similar quantization schemes of lossy quasi-normal modes \cite{van2012spontaneous}, the light-matter interaction is usually described through the dipole approximation, i.e. the relevant electronic transition matrix elements are replaced by point dipole emitters \cite{Scheel2008MQED}.However, the vdW heterostructures studied in the present work, comprise a clear example where the dipole approximation breaks down when emitting lossy optical modes. Furthermore, as shown in this work, the quantum description is essential to capture the deviations from first order order perturbation theory, which we observe when the carrier concentration in the plasmon-carrying graphene sheet becomes sufficiently high.

We have applied the combination of the MQED and DFT to study the coupling of intersubband transitions in atomically thin TMD quantum wells to the optical modes of a doped graphene sheet. Our calculations predict radiative lifetimes below 1 ps and Purcell factors approaching $10^7$. Importantly, these lifetimes are an order of magnitude shorter than the intrinsic low temperature phonon-limited carrier lifetimes in TMDs and thus the spontaneous emission of graphene plasmons from intersubband transitions, should indeed be efficient and observable. We note that Purcell factors as high as the ones reported here are unusual for extended 2D structures, and were previously only reported for geometries confined in all three dimensions\cite{srinivasan2007linear,hennessy2007quantum,loo2012optical}. In contrast to the case of 3D confined cavities where the enhanced emission is due to the coupling to a distinct EM cavity mode, the strong enhancement of the 2D subband transitions considered here is enabled by the significant spatial overlap of the subband wave functions with the highly confined acoustic graphene plasmons.

Although the calculated Purcell factors are indeed very high, they are at least two orders of magnitude smaller than the conventional $Q/V$ enhancement rule highlighting the limitations of the latter. The deviations from the $Q/V$ rule are in part connected to the failure of both the first-order perturbation theory, and the dipole approximation in the examined structures. Our formalism allowed us to benchmark these approximations, which fail in our structure when coupling to high-momentum lossy electron-hole pair excitations in the graphene that increase the total coupling and produce a significant field variation across the TMD quantum well. Other common approximations, such as simpler wave function models can lead to decay rates differing by more than a factor of 10 depending on the details of the model. Apart from the difference in dipole moments, the finer details in the spatial shape of the wave function can affect the coupling to the high-$q$ lossy modes in graphene resulting in a wrong qualitative behavior of e.g. the rate versus graphene doping level.

The developed MQED-DFT method is computationally compact, methodologically simple, and is not limited to a specific system geometry or material type. In addition, the MQED quantization scheme can deal with situations where the emitter is interacting with lossy media, e.g. plasmonic metal structures, doped semiconductors or insulators with polar phonons, which can be included in macroscopic optical parameters of the EM background media. On the other hand, the non-selfconsistent treatment of the light-matter interaction limits the method to the weak to strong coupling regimes where the dressing of the electronic states by the EM field can be neglected. In the ultra-strong coupling regime, the EM field dresses the electronic states, and one must resort to fully self-consistent treatments such as QEDFT. Alternatively, our current theory could be extended using the Power-Zienau-Woolley gauge that overcome several of our approximations when using an electronic multi-polar term \cite{feist2020macroscopic}. However, in the most typical cases of weak-strong coupling, the method developed here provides an efficient and practical alternative to methods like QEDFT, which is applicable to realistic setups in quantum nano-photonics. \\

\acknowledgement
All authors thank Nicholas Rivera for the fruitful discussions. \\

K. S. T. acknowledges support from the Center for Nanostructured Graphene (CNG) under the Danish National Research Foundation (project DNRF103) and from the European Research Council (ERC) under the European Union’s Horizon 2020 research and innovation program (Grant No. 773122, LIMA). \\

I.K. acknowledges the support by the Azrieli Faculty Fellowship, by the GIF Young Scientists’ Program and from the European Research Council (ERC) under the European Union's Horizon 2020 research and innovation  programme under grant Grant No. 851780, and from the Israel Science Foundation grants 3334/19 and 831/19.\\

F.H.L.K. acknowledges financial support from the Government of Catalonia trough the SGR grant, and from the Spanish Ministry of Economy and Competitiveness, through the “Severo Ochoa” Programme for Centres of Excellence in R\&D (SEV-2015-0522), and the Explora Ciencia FIS2017-91599-EXP. F.H.L.K. also acknowledges support by Fundacio Cellex Barcelona, Generalitat de Catalunya through the CERCA program,  and the Mineco grants Plan Nacional (FIS2016-81044-P) and the Agency for Management of University and Research Grants (AGAUR) 2017 SGR 1656.  Furthermore, the research leading to these results has received funding from the European Union’s Horizon 2020  under grant agreement  no. 881603 (Core3) Graphene Flagship, and no. 820378 (Quantum Flagship). This work was supported by the ERC TOPONANOP under grant agreement n$^{\circ}$ 726001.

\subsection*{Author contributions}
F.K., P.S., M.K.S., Y.K., I.K. and K.S.T. were part of the process to conceive the idea for this project. M.K.S. and Y.K. derived and implemented the MQED-DFT framework and K.S.T and I.K. supervised the project. All authors took part in drafting of the manuscript.

\subsection*{Competing interests}
The authors declare no competing interests.

\section{Methods}
\subsection{DFT calculations}
The wave functions of the TMD stacks used to evaluate the light-matter coupling kernel Eq. (\ref{eq:kernel_Ce}) were obtained from DFT employing the PBE exchange-correlation (xc) functional\cite{perdew1996generalized} and a plane wave cut-off of 800 eV. To resolve the subband structure a $\Gamma$-point centered $60\times 60$ $k$-point grid was employed. To avoid spurious interaction between periodically repeated supercells, a 20 Å vacuum was included perpendicular to the TMD stack. The relaxed atomic structures of the four TMD monolayers were obtained from the C2DB database\cite{haastrup2018computational}. Next, $N$-copies of the monolayer were stacked in the 2H-configuration and the interlayer distance was relaxed using the C09-vdW xc-functional\cite{cooper2010van} while keeping the intralayer structure fixed. All calculations were performed with the GPAW code\cite{enkovaara2010electronic}. The graphene sheet and the perfect mirror were both placed at a distance of $(d_\mathrm{Gra} + d_{TMD})/2$ from the outer most atomic layer of the TMD stack, with $d_\mathrm{Gra}$ and $d_\mathrm{TMD}$ being the interlayer distances in graphene and the TMD respectively.

\subsection{1D wave function approximations}
In addition to the full 3D DFT generated wave functions, this work considers two different wave function approximations: Particle in a box states, and the in-plane average of the DFT states. The particle in a box states are given as,

\begin{align}
    \phi_n(\boldsymbol r) = \left(\frac{2}{d}\right)^{1/2}\mathrm{sin}\left(\frac{n\pi z}{d}\right)\,,
\end{align}

\noindent where $n$ is the subband index and $d$ is the width of the TMD layer. The 1D DFT states are found when approximating $\phi_n(\boldsymbol r)$ with the in-plane average of the full 3D wave functions.

\subsection{The Dyadic Green's function}
In this section we will find the Dyadic Green’s function(DGF) for the structure discussed in the main text. The DGF, $\boldsymbol{\tensor{G}}(\boldsymbol r, \boldsymbol r', \omega)$, provides a general description of the electromagnetic environment of a structure, and it plays a fundamental role in determining the interaction between matter and the modes of the electromagnetic field. $\boldsymbol{\tensor{G}}(\boldsymbol{r}, \boldsymbol{r'}, \omega)$ is defined as propagating a current density, $\boldsymbol{j}(\boldsymbol{r'}, \omega)$, at a source points $\boldsymbol{r'}$, onto the corresponding electric field at point $\boldsymbol{r}$ according to:

\begin{align}\label{eq:E_field_from_dipole_and_DGF}
    \boldsymbol{E}(\boldsymbol{r}) = i\omega \mu_0\mu(\boldsymbol{r})\int_V \boldsymbol{\tensor{G}}\left(\boldsymbol{r}, \boldsymbol{r'}\right)\boldsymbol{j}(\boldsymbol{r'})dV'
\end{align}

\noindent This translates into the DGF itself being defined as the electric field $\boldsymbol{E}$ at point $\boldsymbol{r}$, generated by a radiating dipole, $\boldsymbol{p}$, located at the source point $\boldsymbol{r'}$. Therefore, the field from the radiating dipole can be expressed as, $\boldsymbol{E}(\boldsymbol{r},\omega)=\omega^2\mu_0\mu\boldsymbol{\tensor{G}}(\boldsymbol{r},\boldsymbol{r'}, \omega)\cdot\boldsymbol{p}$\cite{novotny2012principles}. Specifying to non-magnetic materials, $\mu = 1$, the equation of motion for the DGF is the Helmholtz equation with a point source term:

\begin{align}
    \left[\boldsymbol{\nabla}\times \boldsymbol{\nabla}\times - \left(\frac{\omega^2}{c^2}\right) \boldsymbol{\tensor{\epsilon}}(\boldsymbol{r}, \omega) \right] \boldsymbol{\tensor{G}}(\boldsymbol{r}, \boldsymbol{r'}, \omega) = \delta(\boldsymbol{r} -\boldsymbol{r'})
\end{align}

\noindent where $\omega$ is the angular frequency and $c$ is the speed of light. In this work, we consider the electric permittivity tensor, $\boldsymbol{\tensor{\epsilon}} = \mathrm{diag}\left(\epsilon_\parallel, \epsilon_\parallel, \epsilon_\bot \right)$ to account for anisotropic dielectric properties of the TMD stacks. \newline
Because of the layered nature of the geometry, the in-plane translational invariance of the system enables the DGF to be expressed as a mode expansion over the in-plane momentum, $\boldsymbol{q} = (q_x, q_y)$ in the following way\cite{chew1995waves,novotny2012principles,wasey2000efficiency}:

\begin{align}
    \boldsymbol{\tensor{G}}\left(\boldsymbol{r}, \boldsymbol{r'}, \omega\right) = \int d\boldsymbol{q}e^{i\boldsymbol{q}\cdot(\boldsymbol{\rho} - \boldsymbol{\rho'})}\left[\boldsymbol{\tensor{N}}(\boldsymbol{q}, z, z') + \boldsymbol{\tensor{M}}(\boldsymbol{q}, z, z'))\right]
\end{align}

\noindent where $\boldsymbol{\tensor{N}}(\boldsymbol{q}, \boldsymbol{z}, \boldsymbol{z'})$ and $\boldsymbol{\tensor{M}}(\boldsymbol{q}, \boldsymbol{z}, \boldsymbol{z'})$ describes the spectral response of the TE and TM polarizations respectively. However, since the cavity plasmon mode is TM polarized, we retain only the TM polarized part of the DGF. 

To further derive the explicit form of the Green's function in the multilayered structure, we use express $\boldsymbol{\tensor{M}}(\boldsymbol{q}, \boldsymbol{z}, \boldsymbol{z'})$ with the reflection coefficients from each side of the TMD \cite{chew1995waves,wasey2000efficiency}. It can be shown that the components of the TM response are
\begin{align}\label{eq:Im_M}
    & \mathrm{Im}\left[M_{\boldsymbol{\hat{q}}\boldsymbol{\hat{q}}}(\boldsymbol{q}, z, z')\right]= \frac{k_zc^2}{4\pi^2\omega^2\epsilon_\parallel}\mathrm{sinh}(k_z z)\mathrm{sinh}(k_z z') \mathrm{Im}\left[\frac{r_g e^{-2k_zd} - 1}{r_g e^{-2k_zd} + 1}\right] \\
    & \mathrm{Im}\left[M_{\boldsymbol{\hat{z}}\boldsymbol{\hat{q}}}(\boldsymbol{q}, z, z')\right]= \frac{iqc^2}{4\pi^2\omega^2\epsilon_\bot}\mathrm{cosh}(k_z z)\mathrm{sinh}(k_z z') \mathrm{Im}\left[\frac{r_g e^{-2k_zd} - 1}{r_g e^{-2k_zd} + 1}\right] \\
    & \mathrm{Im}\left[M_{\boldsymbol{\hat{q}}\boldsymbol{\hat{z}}}(\boldsymbol{q}, z, z')\right]= \frac{-iqc^2}{4\pi^2\omega^2\epsilon_\bot}\mathrm{sinh}(k_z z)\mathrm{cosh}(k_z z') \mathrm{Im}\left[\frac{r_g e^{-2k_zd} - 1}{r_g e^{-2k_zd} + 1}\right] \\
    & \mathrm{Im}\left[M_{\boldsymbol{\hat{z}}\boldsymbol{\hat{z}}}(\boldsymbol{q}, z, z')\right]= \frac{q^2c^2\epsilon_\parallel}{4\pi^2\omega^2k_z\epsilon_\bot^2}\mathrm{cosh}(k_z z)\mathrm{cosh}(k_z z') \mathrm{Im}\left[\frac{r_g e^{-2k_zd} - 1}{r_g e^{-2k_zd} + 1}\right]
\end{align}

\noindent where $k_z = \sqrt{\frac{\epsilon_\parallel}{\epsilon_\bot}q^2 - \epsilon_\parallel\frac{\omega^2}{c^2}}$ is the wave vector in the out of plane direction, and $\boldsymbol{\hat{q}}$ is an in plane spatial unit vector pointing in the direction of $\boldsymbol{q}$. In addition, we have assumed perfect reflection from the TMD to the metal and used the quasistatic reflection coefficient $r_g = \frac{\sqrt{\epsilon_\parallel\epsilon_\bot} - 1 - i \frac{q\sigma_s(q, \omega)}{\epsilon_0\omega}}{\sqrt{\epsilon_\parallel\epsilon_\bot} + 1 + i \frac{q\sigma_s(q, \omega)}{\epsilon_0\omega}}$ as the TMD to Graphene reflection coefficient (see Supplementary note 3 for derivation) with $\sigma_s(q, \omega)$ as the nonlocal graphene surface conductivity which includes plasmonic losses and Landau damping\cite{gonccalves2016}. We note that the Green's function expressions can be derived directly from the electric fields from a dipole oriented in the $\boldsymbol{\hat{q}}$ and  $\boldsymbol{\hat{z}}$ directions. Finally, for our further calculations we will also use the identity, $\mathrm{Im}\left[\boldsymbol{\tensor{G}}\left(\boldsymbol{r}, \boldsymbol{r'}, \omega\right)\right] = \int d\boldsymbol{q}e^{i\boldsymbol{q}\cdot(\boldsymbol{\rho} - \boldsymbol{\rho'})}\mathrm{Im}\left[\boldsymbol{\tensor{M}}(\boldsymbol{q}, z, z'))\right]$, which describes the entire TM optical response that the TMD experiences, whether we want to consider the TMD emitter as local, nonlocal or even a dipole emitter.

\subsection{Background permittivity}
\paragraph{TMD permitivity}

The permitivities of the TMD's were calculated both in-plane and out-of-plane using the GPAW code\cite{enkovaara2010electronic} and they are consistent with the values found in the literature\cite{laturia2018dielectric}. The permittivities are chosen not to include the intersubband transitions as to avoid double counting and are thus frequency independent. The values used are shown in table \ref{tab:TMD_perm}.

\begin{table}[ht]
\begin{tabular}{|l|l|l|}
\hline
Material & In-plane static dielectric constant & Out-of-plane dielectric constant \\ \hline
MoS$_2$  & 16.1                                & 7.2                              \\ \hline
WS$_2$   & 14.5                                & 6.6                              \\ \hline
MoSe$_2$ & 17.3                                & 8.3                              \\ \hline
WSe$_2$  & 15.7                                & 7.8                              \\ \hline
\end{tabular}
\caption{The TMD material permittivities that were used in our simulations.}
\label{tab:TMD_perm}
\end{table}

\paragraph{Nonlocal graphene conductivity}

\begin{figure}[!htb]
    \centering
    \includegraphics[width=0.5\textwidth]{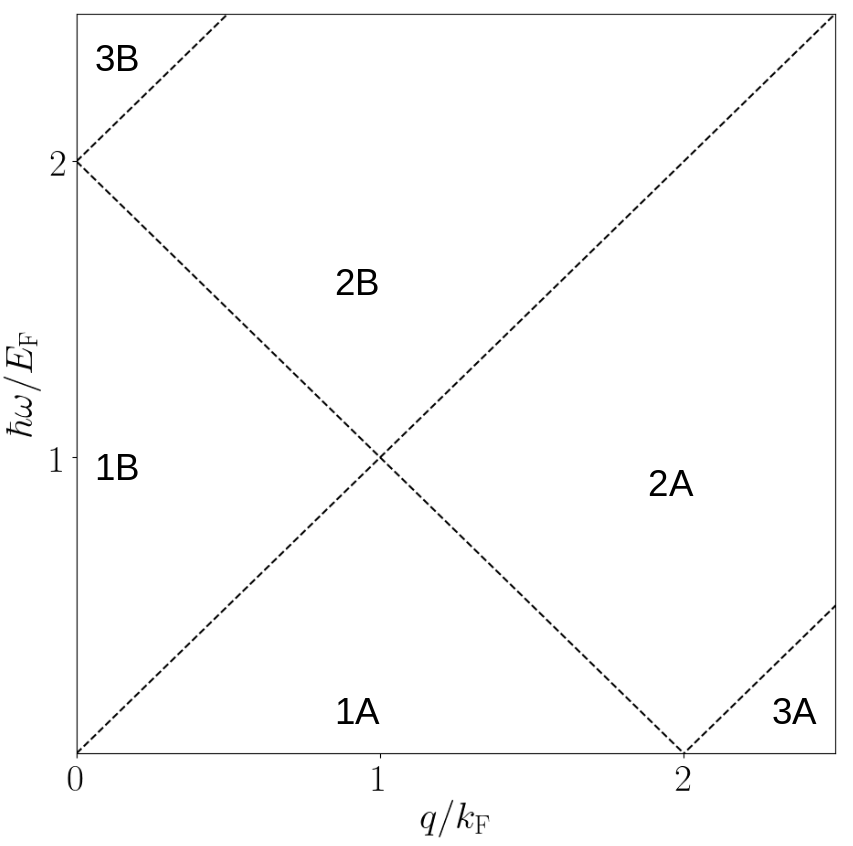}
    \caption{Regions on which the graphene conductivity is to be calculated.}
    \label{fig:graphene_regions}
\end{figure}

In order for the light-matter interactions to take into account all possible optical excitations in the graphene, including the high-momentum e-h continuum, there is a necessity for a non-local graphene conductivity as opposed to a simple Drude model. This approach include any possible optical response in the  graphene, within the random phase approximation. In this chapter, we will explain how to retrieve the non-local graphene surface condictivity in a similar manner as found in Peres and Goncalves \cite{gonccalves2016}.We note that treating the graphene as a 3D material with a finite thickness (with a bulk conductivity instead of a surface conductivity) should not affect the graphene plasmonic properties.
First, the graphene conductivity, $\sigma(q, \omega)$,  is defined in the following way:

\begin{align}
    \sigma(q, \omega) = \frac{4i\sigma_0\hbar\omega}{q^2}\chi_t\left(\frac{q}{k_F}, \frac{\hbar \omega}{E_F}\right)
\end{align}

\noindent with $\sigma_0=\frac{e^2}{4\hbar}$, and $\chi_t$ being the susceptibility of normalized variables, defined as:

\begin{align}
    \chi_t(x, y) = \frac{\left(1 + i\frac{\Gamma}{yE_F}\right)\chi_g(x, y+i\frac{\Gamma}{E_F})}{1 + i\frac{\Gamma}{yE_F}\chi_g(x, y+i\frac{\Gamma}{E_F})/\chi_g(x, 0)}
\end{align}

\noindent with $\Gamma$ as the graphene dampening parameter. This work uses the value of $\Gamma$ to be 16 meV which is consistent with choice by Dias \textit{et. al.}\cite{dias2018probing}. Finally $\chi_g(x, y)$ is defined according to the regions in figure \ref{fig:graphene_regions}, that describe the regions of different optical responses in the graphene. For more readable expressions, we define the following auxiliary functions, $F(\xi) = \xi\sqrt{\xi^2 - 1} - \textrm{arccosh}(\xi)$, and, $C(\xi) = x\sqrt{1- \xi^2} - \textrm{arccos}(\xi)$. Furthermore, $x = q/k_\mathrm{F}$ is the normalized momentum and $y = \hbar \omega/E_\mathrm{F}$ is the normalized energy.

The different regions of the responses are:
\paragraph{Region 1A:} Region 1A is defined as the part of the $(q, \omega)$ plane where $x > \mathrm{Re}[y]$ and $\mathrm{Re}[y] + x < 2$. In this region the graphene electrons stay in the same band while changing largely their momentum. The expression for the nonlocal susceptibility is:

\begin{align}
    \chi^{1A}_g(x, y) = \frac{-2}{\pi}\frac{k_\mathrm{F}}{\hbar v_\mathrm{F}} + i\frac{k_\mathrm{F}}{\hbar v_\mathrm{F}}\frac{x^2}{4\pi\sqrt{x^2 - y^2}}\left[F\left(\frac{2 - y}{x}\right) - F\left(\frac{2 + y}{x}\right)\right]
\end{align}

\paragraph{Region 2A:} Region 2A is defined by $x-\mathrm{Re}[y] < 2$, $x > \mathrm{Re}[y]$ and $x + \mathrm{Re}[y] > 2$. In this region the graphene electrons change their band while also changing largely their momentum. The expression for the susceptibility is:

\begin{align}
    \chi^{2A}_g(x, y) & = \frac{-2}{\pi}\frac{k_\mathrm{F}}{\hbar v_\mathrm{F}} - \frac{k_\mathrm{F}}{\hbar v_\mathrm{F}}\frac{x^2}{4\pi\sqrt{x^2 - y^2}}C\left(\frac{y - 2}{x}\right) \nonumber \\
    & -i\frac{k_\mathrm{F}}{\hbar v_\mathrm{F}}\frac{x^2}{4\pi\sqrt{x^2 - y^2}}F\left(\frac{2+y}{x}\right)
\end{align}

\paragraph{Region 3A:} Region 3A is defined by $x-2 > \mathrm{Re}[y]$. In this region there is no direct transition for the graphene and thus the optical response is weak. The expression for the susceptibility is:

\begin{align}
    \chi_g^{3A}(x, y) = \frac{-2}{\pi}\frac{k_\mathrm{F}}{\hbar v_\mathrm{F}} + \frac{k_\mathrm{F}}{\hbar v_\mathrm{F}}\frac{x^2}{4\pi\sqrt{x^2 - y^2}}\left[C\left(\frac{2 + y}{x}\right) - C\left(\frac{y - 2}{x}\right)\right]
\end{align}

\paragraph{Region 1B:} Region 1B is defined by $\mathrm{Re}[y] > x$ and $\mathrm{Re}[y] + x < 2$. This region enables the propagating graphene plasmons while inhibiting other e-h excitations.The expression for the susceptibility is:

\begin{align}
    \chi^{1B}_g(x, y) = \frac{-2}{\pi}\frac{k_\mathrm{F}}{\hbar v_\mathrm{F}} + \frac{k_\mathrm{F}}{\hbar v_\mathrm{F}}\frac{x^2}{4\pi\sqrt{y^2 - x^2}}\left[F\left(\frac{2 + y}{x}\right) - F\left(\frac{2 - y}{x}\right)\right]
\end{align}

\paragraph{Region 2B:} Region 2B is defined by $\mathrm{Re}[y] > x$, $\mathrm{Re}[y] + x > 2$ and $\mathrm{Re}[y] - x < 2$. This region also enables the propagating graphene plasmons but includes other e-h excitations. The expression for the susceptibility is:

\begin{align}
    \chi_g^{2B}(x, y) & = \frac{-2}{\pi}\frac{k_\mathrm{F}}{\hbar v_\mathrm{F}} + \frac{k_\mathrm{F}}{\hbar v_\mathrm{F}}\frac{x^2}{4\pi\sqrt{y^2 - x^2}}F\left(\frac{2 + y}{x}\right) \nonumber \\
    & + \frac{k_\mathrm{F}}{\hbar v_\mathrm{F}}\frac{x^2}{4\pi\sqrt{y^2 - x^2}}C\left(\frac{2 - y}{x}\right)
\end{align}
\paragraph{Region 3B:} Region 3B is defined as $\mathrm{Re}[y] - x > 2$ and it captures all of the high energy interband transitions. The expression for the susceptibility is:

\begin{align}
    \chi_g^{3B}(x, y) & = \frac{-2}{\pi}\frac{k_\mathrm{F}}{\hbar v_\mathrm{F}} + \frac{k_\mathrm{F}}{\hbar v_\mathrm{F}}\frac{x^2}{4\pi\sqrt{y^2 - x^2}}\left[F\left(\frac{2 + y}{x}\right) - F\left(\frac{2 - y}{x}\right)\right] \nonumber \\
    &-i\frac{k_\mathrm{F}}{\hbar v_\mathrm{F}}\frac{x^2}{4\sqrt{y^2 - x^2}}
\end{align}

\suppinfo

\section{Data availability}
The data that support the plots and findings of this paper are available from M.K.S. (markas@dtu.dk) upon reasonable request.



\end{document}